\documentclass[11pt,en]{elegantpaper}

\title{Network topology of the Euro Area interbank market}
\author{
  Ilias Aarab
  \and
  Thomas Gottron
}
\institute{
  European Central Bank\footnote{The views expressed in this paper are those of the authors and do not necessarily reflect those of the European Central Bank.},
  Sonnemannstraße 20, Frankfurt am Main, Germany\\
  \email{Ilias.Aarab@ecb.europa.eu}\\
  \email{Thomas.Gottron@ecb.europa.eu}
}
\date{20/10/2022}

\usepackage{url}

\usepackage{amsmath}
\usepackage{threeparttable}
\usepackage{adjustbox}
\usepackage{comment}
\usepackage{longtable,booktabs}
\usepackage{xcolor}

\usepackage{etoolbox}
\makeatletter
\patchcmd\longtable{\par}{\if@noskipsec\mbox{}\fi\par}{}{}
\makeatother

\IfFileExists{footnotehyper.sty}{\usepackage{footnotehyper}}{\usepackage{footnote}}
\makesavenoteenv{longtable}

\begin{document}

\maketitle

\begin{abstract} % Ilias: max 150 words
The rapidly increasing availability of large amounts of granular financial data, paired with the advances of big data related technologies induces the need of suitable analytics that can represent and extract meaningful information from such data.
\begin{comment}
Furthermore, economic downturns like the Great Financial Crisis have raised the need for an in-depth comprehension of the topologies of the various financial systems that are deeply intertwined with each other. Such systems are predominantly driven by the interconnectedness of financial entities through multi-faceted and complex far-reaching relationships. It is therefore essential to better understand the topological structures of such systems in order to discern these embedded relationships and their potential impact on the system as a whole.
\end{comment}
In this paper we propose a multi-layer network approach to distill the Euro Area (EA) banking system in different distinct layers. Each layer of the network represents a specific type of financial relationship between banks, based on various sources of EA granular data collections. The resulting multi-layer network allows one to describe, analyze and compare the topology and structure of EA banks from different perspectives, eventually yielding a more complete picture of the financial market. This granular information representation has the potential to enable researchers and practitioners to better apprehend financial system dynamics as well as to support financial policies to manage and monitor financial risk from a more holistic point of view.
\keywords{granular data collections, network topology, interbank system, multi-layer networks}
\end{abstract}

% Ilias: Introduction is allowed to begin on abstract page. We should merge Intro and Related work into one section.
\section{Introduction}\label{sec:intro}

The approach of modelling the banking sector as a network of interconnected entities has received increasing attention in recent years~\cite{huser2015too}. The increased attention is driven by the demand to better understand the complexity of interrelations between banks after the 2008 financial crisis, where mechanisms of an interwoven financial system were difficult to understand and predict. Consequently, more and more granular data\footnote{In the given context, granular data refers to detailed, specific, and finely segmented information about individual components or entities within the financial and banking system. This data provides a detailed view of individual banks, financial institutions, or other entities, including their transactions, interactions, and various aspects of their operations.} about the financial and banking system is made available to supervisors to fuel analytical models capturing interdependencies in the market.

Related literature can be divided into two categories: (1) motivating work, arguing for the advantages of modelling financial markets as networks, and (2) empiric work, using different types of data to describe certain aspects of the network structures in the banking sector.

As argued by \cite{allen2009networks}, the use of network theory can help in understanding financial systems, the interactions among agents and certain economic phenomena.
According to \cite{bandt2012systemic}, the complex network of exposures among banks can be considered as one of the key features for determining systemic risk while \cite{di2018survey} review network-based approaches as analytical basis for systemic risk indicators.
Recent work focuses in particular on multi-layer networks~\cite{bargigli2015multiplex}. %,aldasoro2018multiplex}.
Multi-layer networks cater for the different types of interrelation among banks and model them separately in distinct layers of the network, where each layer may exhibit a different topology~\cite{battiston2014structural}.%\cite{,langfield2016interbank}

Empirical work analysing interbank networks is mainly driven by the availability of data. Relevant papers range from descriptive approaches to investigating the network topology to applying novel analytical frameworks to the data.
\cite{boss2004network} analysed Austrian banks based on general network theory methods and showed that the degree distribution in the network follows a powerlaw distribution, that the network has a low clustering coefficient and a short average path length.
Key insights regarding the topology of an emerging network are that a core-periphery structure describes the network better than other generative models~\cite{van2014finding}. % like preferential attachment \cite{barabasi1999emergence}.
%Similar structural observations are made also on a network of German banks and using data of bilateral exposures~\cite{craig2014interbank}.
%The authors explain this phenomenon by the banks’ specialisations as a reason that networks do not form randomly.
%An analysis of large EU banks~\cite{alves2013structure} distinguishes between different types of interconnections, in particular also between direct and indirect ways and considers temporal effects and dynamics in the networks.
%
Empirical investigations of multi-layer networks capture different types of connections between banks. \cite{bargigli2015multiplex} distinguish between maturity and the secured or unsecured nature of the contracts among banks on the Italian market and showed how the resulting layers are of quite different topology.
Moreover, a flattened and aggregated representation of all layers is not suitable to capture the full complexity of the network. \cite{montagna2016multi} develop their model for assessing systemic risk on a multi-layer network of short-term and long-term loans and the common exposures to financial assets due to overlapping portfolios. %, partially leveraging real world data and filling the gaps with estimates.
\cite{aldasoro2018multiplex} use data on exposures between large European banks, broken down by  maturity and instrument type to characterise the main features of a multi-layer network. %In contrast our multi-layer network is driven by granular data collections and represents a network of all Euro area Significant Institutions.

In this paper, we highlight the utility of granular data collections for constructing and analyzing multi-layer networks. Granular data offers flexibility in filtering, aggregation, and focus, enabling the construction of diverse network models from the same source data. Additionally, specific instances of networks can serve multiple analytical purposes and answer various questions. We demonstrate the construction of a multi-layer network covering the broad scope of the EA financial market using data from different granular collections, detailing the preprocessing, integration, and aggregation methods applied. We present analytical tasks, implemented approaches, and insights gained from the analysis, aligning real-world data results with theoretical expectations.

\section{Data}\label{sec:data}

Diverse granular datasets serve as the fundamental basis for constructing our multi-layer networks. In the subsequent sections, we explain their distinct characteristics and provide a rationale for their selection in our analysis:

\textbf{Register of Institutions and Affiliates Data (RIAD):} RIAD is the shared master dataset serving several European System of Central Banks (ESCB) and Single Supervisory Mechanism (SSM) business processes and statistical data collections. The RIAD data model includes more than 100 properties for legal entities, including a wide range of entity identifiers and relationships (e.g. foreign branches, subsidiaries).
In our application, RIAD served to identify entities across different datasets and as source for constructing banking groups.

\textbf{List of significant banking groups (ROSSI list):}
All of the granular datasets considered in this work comprise various information on the largest banking groups in the EA and participating member states which are directly supervised by the SSM. Therefore, the ROSSI list is used to determine the sample of banking groups to be included in our multi-layer network.
As of 1\textsuperscript{st} of June 2021, this list of significant institutions comprised a total of 114 banking groups.
From ROSSI, we extract the RIAD codes of significant institutions, considering the prudential consolidation regime. This regime widens the group structure view as it also considers cases where the consolidating entity might be different than a bank (e.g. financial holding) and improves consistency in the aggregation of granular datasets.

\textbf{CSDB and SHSG data:} The Centralised Securities Database (CSDB) %/cite{european2012financial}
is a reference database for all securities relevant for statistical purposes of the ESCB. It comprises information on debt securities, equity instruments and investment fund shares which are stored on a security-by-security basis. Each instrument is identifiable by its International Securities Identification Number.
While CSDB covers the issuance of securities by banking groups, the Securities Holdings Statistics Database by Group (SHSG) dataset contains the holdings of securities by banks. Thus, using both databases we gain insights into the securities being issued by banking groups as well as the securities they are currently holding.
%, inside and outside the euro area.

\textbf{AnaCredit data:} The AnaCredit dataset %~\cite{israel2017analytical}
contains loan-by-loan information on loans extended to corporations collected from EA banks. This data is reported at monthly frequency starting from September 2018. Among others, information on the outstanding amount, maturity, interest rate, collateral/guarantee, and on involved counterparties is collected for each of the individual loans.
%This makes AnaCredit an incredibly rich dataset that can be used for various analytical purposes.

\textbf{SFT – Securities Financing Transactions data}: The Securities Financing Transactions Data Store (SFTDS) collects and processes data reported under the Securities Financing Transactions Regulation (EU) 2015/2365\footnote{Regulation (EU) No. 2015/2365 on transparency of securities financing transactions and of reuse and amending Regulation (EU) No. 648/2012 Securities Financing Transactions Regulation – SFTR).}.
SFTs and the scope of the SFTR include\footnote{Article 3 (11) SFTR}:
(a) repurchase/ reverse repurchase transactions, (b) buy-sell back or sell-buy back transactions, (c) securities or commodities lending and securities or commodities borrowing and (d) margin lending transactions.

\textbf{FINREP/COREP:} FINREP (Financial Reporting Standards) and COREP (Common Reporting Standards) are the two reporting frameworks developed as part of the implementation of Basel III in Europe\footnote{EBA’s Implementing Technical Standards (ITS) amending the European Commission’s Implementing Regulation (EU) No 680/2014 on supervisory reporting of institutions under Regulation (EU) No 575/2013.} which gives the European Banking Authority the legal basis to request both capital and financial information from EA banks.
COREP specifies the framework related to capital information, while FINREP specifies the financial information.
Banks subject to International Financial Reporting Standards (IFRS) already use the FINREP templates to submit financial information in a harmonised format at consolidated level. This requirement has also been extended to SSM banks reporting at sub-consolidated or solo level under IFRS and national Generally Accepted Accounting Principles (GAAPs)\footnote{Regulation (EU) 2015/534 of the ECB of 17 March 2015}. FINREP templates are subject to a quarterly mandatory reporting.
COREP is used to collect Pillar 1 data and information on liquidity, leverage and large exposures from banks in a harmonised format.

\section{Network modelling}\label{sec:modelling}

\subsection{Identification and modelling of banking groups}\label{mod1}

The first step in modelling a multi-layer network is to define the nodes $n_i$ of the network. As our main focus is to model the exposures between banking groups across different markets, we model banking groups as nodes. The construction of banking groups is initiated from the head of a banking group. The list of group heads of significant\footnote{The definition of Significant Institutions can be found in Regulation (EU) No 468/2014 of the European Central Bank of 16 April 2014.} banks in the EA was retrieved from ROSSI.
Upon identifying the group head, we discern all entities affiliated with the respective group. A subsidiary of a group is identified as an entity on which the group head exercises direct or indirect control, based on equity share\footnote{In addition, supervised entities according to Guideline (EU) 2020/497 of the ECB are also included in the group, if not already included following the control relationship criteria.}.
To obtain a complete group structure, including subsidiaries that are not credit or financial institutions, we leverage and integrate different group information available in ROSSI, COREP and RIAD.

\subsection{Granular data integration}\label{mod2}

After the identification of banking groups and their subsidiaries, we integrate the different granular datasets as basis for forming the interactions between groups.
Consistent identification of group entities is realised using entity identifiers available in RIAD.
During the integration process we also harmonise and align different structures and semantics of the various datasets (e.g., stock-based vs. transactional data).
Having information on the banking group structure allows us to aggregate information on subsidiaries to the group head level and identify interrelations across all granular datasets\footnote{To efficiently process the large amount of data we make use of a Hadoop based data lake to store the data and utilize Spark for distributed processing.}.

\subsection{Construction of a multi-layer network}\label{mod3}

%Based on the list of banking groups and after integrating and aggregating the granular data, we construct the multi-layer network.
We follow~\cite{battiston2014structural} and define a multi-layer network as a network where each node appears in a set of different layers $\ell$, and each layer describes all the edges of a given type. In our case, the edges $e_{ij}^\ell$ from node $n_i$ to node $n_j$ are directed and their weight indicates the aggregated exposure of banking group $n_i$ towards $n_j$ within layer $\ell$\footnote{Note that intra-group exposures of banking groups are removed to avoid self-loops within the network.}. The edge values differ across the layers $\ell$ based on the type of exposure they represent:

\textbf{Long-term credit layer}: This layer contains loan exposures between banking groups having an initial maturity of at least three months, in line with~\cite{montagna2016multi}. Exposures are aggregated over different types of loan instruments (e.g., deposits, credit lines, convenience credit, etc.), with the bulk (more than 90\%) of exposures coming from deposits, credit lines other than revolving credit and unspecified loans. Edge weights $e_{ij}^{\textit{ltc}}$ indicate the outstanding nominal values.

\textbf{Short-term credit layer}: This layer contains loan exposures with an initial maturity shorter than three months. Exposures are processed in the same way as for the long-term credit layer, with the additional constraint of ensuring strictly positive maturities of the loans. We observe that a large part of instruments (more than 45\%) are reverse repurchase agreements confirming the short-term nature of the layer. Edge weights $e_{ij}^{\textit{stc}}$ indicate the outstanding nominal values.

\textbf{Cross-securities layer}: This layer contains equity and debt holdings between banking groups. An edge from node $n_i$ to node $n_j$ represents the market value of the securities issued by $n_j$ and held by $n_i$. We do not distinguish between equity and debt-based securities and aggregate them together into the same edge. The edge weight $e_{ij}^{\textit{cs}}$ represents the market value of investments made by $n_i$ into $n_j$’s issued securities.

\textbf{Short-term funding layer}: The short-term funding layer contains repurchasing agreements and buy-sellback information obtained from the SFT dataset. Within this layer an edge from node $n_i$ to $n_j$ represents the (aggregated) open funding transactions between banking group $n_i$ and $n_j$. The edge $e_{ij}^{\textit{stf}}$ is directed from the collateral taker to the collateral giver. To integrate the resulting data with the other granular datasets, we extract those transactions that are indicated as active at the end of a month as they represent a snapshot of the end-of-month exposures of banking groups.

\textbf{Overlapping portfolio layer}: The overlapping portfolio layer models the investments made by banking groups in securities that are issued by entities that are not part of the interbank market. We are particularly interested in the way that this layer can induce potential fire sales effects. To model this behaviour in a straightforward manner, we create an undirected graph where an edge $e_{ij}^{op}$ between banking group $n_i$ and banking group $n_j$ represents the market value of the overlapping part of their portfolios. In this manner, the effect of a sell-off of $n_i$’s securities, can be traced back by following the edges that are (directly or indirectly) connected with $n_i$. To contain the universe of possible securities, we aggregate them on the issuer level. This approach offers the added advantage of implicitly accounting for the positive correlations among securities issued by the same issuer.

\textbf{Flattened network layer}: The flattened network layer is an artificially constructed layer, in the sense that it does not represent a specific real-world market. Instead, the layer is defined as the aggregation of all of the layers of the multi-layer network. More specific, we follow~\cite{battiston2014structural} and create a weighted aggregated overlapping
adjacency matrix to form the flattened network layer.  %The sum between two layers being defined as the aggregation of all the edges in the layers. In case the same edge from banking group i to j is present in both layers, their weights are added up. In case an edge from i to j is present in one layer, but not the other, it is simply added to the aggregated layer. The same logic applies when summing more than two layers together. Care needs to be taken however when adding undirected layers, in our case the external securities layer.

\subsection{Enrichment of the nodes}\label{mod4}

After constructing the multi-layer network, we further enrich the nodes of the network.
The goal is to describe the nodes through their most relevant balance sheet items like the Tier 1 Capital and Total Assets. These balance sheet values are retrieved from FINREP\footnote{Total assets are available in FINREP template: F 01.01 - Balance Sheet Statement.} and COREP\footnote{Tier 1 Capital is available in the COREP template: C47.00 – Leverage ratio calculation. The amount of Tier 1 capital is calculated according to article 25 of the CRR, without taking into account the derogation laid down in Chapters 1 and 2 of Title I of Part Ten of the CRR.} templates, where the values are available at a consolidated level allowing for a mapping to the banking groups via RIAD identifiers.\footnote{Note that the analysis we conduct focuses on a static snapshot of the multi-layer network. The considered snapshot is the end-of-month observation of the network as of June 2021. This is a date in which all datasets, considering their respective frequencies, are available.}

\section{Network topology}\label{sec:network_topology}

To dig deeper into the topology of the constructed multi-layer network, we leverage both graph statistics, which describe the overall network structure, and centrality measures aimed at characterizing individual nodes within the system. As seen in related work \cite{bargigli2015multiplex}, this is of particular interest, as the different semantics of the layers in a multi-layer network entail different topologies and structures. The use of graph statistics jointly with centrality measures is necessary, as congruence of topology between two or more layers (graph statistics) does not necessarily imply point-wise similarity between the nodes in the layers (centrality measures)~\cite{Engel2021}.

\subsection{Graph statistics}

Table~\ref{tab1} presents an overview of the global graph metrics across the different layers. The multi-layer network encompasses $114$ banking groups interacting across its various layers. Notably, the overlapping portfolio layer exhibits significant interconnectedness, evidenced by 3,614 edges. In contrast, both the short-term and long-term interbank credit layers, along with the short-term funding layer, display sparser connectivity, with densities ranging from 4\% to 7\%. The cross-securities layer falls between these extremes, featuring 2,456 interlinkages among the agents.\footnote{To have a meaningful comparison between the layers, we transform the undirected overlapping portfolio layer into a directed symmetric graph.}

\begin{table}[t]
\centering
\caption{Graph statistics across network layers}
\label{tab1}
\small
\setlength{\tabcolsep}{4pt}

\begin{adjustbox}{width=\linewidth}
\begin{tabular}{lrrrrrrr}
\toprule
& \multicolumn{1}{c}{ST credit}
& \multicolumn{1}{c}{LT credit}
& \multicolumn{1}{c}{Cross-sec.}
& \multicolumn{1}{c}{ST funding}
& \multicolumn{1}{c}{Overlap}
& \multicolumn{1}{c}{Overlap (dir.)}
& \multicolumn{1}{c}{Flattened} \\
\midrule
Nodes ($N$)                       & 114 & 114 & 114 & 114 & 114 & 114 & 114 \\
Edges ($E$)                       & 525 & 901 & 2{,}456 & 900 & 3{,}614 & 3{,}614 & 2{,}969 \\
Connected components              & 26  & 14  & 11  & 42  & 13  & 13  & 4 \\
Largest component: share of nodes & 0.78 & 0.88 & 0.91 & 0.64 & 0.89 & 0.89 & 0.97 \\
Largest component: share of edges & 0.79 & 0.76 & 0.77 & 0.65 & 1.00 & 1.00 & 0.71 \\
Diameter (largest component)      & 5   & 4   & 5   & 4   & 2   & 2   & 4 \\
Avg. clustering coefficient       & 0.29 & 0.38 & 0.51 & 0.41 & 0.79 & 0.39 & 0.62 \\
Reciprocity                       & 0.42 & 0.48 & 0.47 & 0.71 & 0.00 & 0.00 & 0.58 \\
Density                           & 0.04 & 0.07 & 0.19 & 0.07 & 0.56 & 0.28 & 0.23 \\
Global efficiency                 & 0.30 & 0.40 & 0.55 & 0.24 & 0.68 & 0.68 & 0.63 \\
Herfindahl index                  & 0.05 & 0.04 & 0.02 & 0.04 & 0.07 & 0.07 & 0.04 \\
\bottomrule
\end{tabular}
\end{adjustbox}

\vspace{2pt}
\begin{minipage}{\linewidth}
\footnotesize
\emph{Notes:} ST = short-term; LT = long-term; Cross-sec. = cross-securities; dir. = directed (symmetrized) version of the overlapping-portfolio layer.
\end{minipage}
\end{table}

Although no layer is fully connected, a significant majority of nodes belong to the largest connected component~\footnote{A connected component is a set of nodes which are directly or indirectly connected to each other. From the perspective of a specific node, it is not possible to reach any node outside the component it belongs to. } within each layer. Specifically, in the cross-securities and credit layers, approximately 80-90\% of nodes are part of the largest subcomponent. In the short-term funding layer, this proportion reduces to 64\% of agents forming the largest component. Notably, in the overlapping portfolio layer, 90\% of nodes constitute the largest component; all realized edges are within this component, suggesting that the remaining agents exist as isolated islands.

\begin{comment}
Zooming into the fragmentations of the different layers, we observe that
no layer is fully connected. Especially the short-term funding layer
seems to be highly fragmented, with 42 distinct subcomponents detected
within the graph. Contrary to the funding layer, other layers seem to be
more connected, with the number of subcomponents ranging from 11
(cross-securities) to 14 (long-term credit), with the short-term credit
layer being in between with 26 subcomponents.

However, this picture changes when analyzing the largest subcomponent of
the different layers, where we see that most of the nodes are part of
the largest subcomponent within each layer. More precise, for the
cross-securities and credit layers we see that about 80-90\% of the
nodes are part of the largest subcomponent. For the short-term funding
layer, we see that only 64\% of the agents make part of the largest
component. Focusing on the overlapping portfolio layer, we see that,
though 90\% of the nodes are part of the largest component, all of the
realized edges are within this component, indicating that the remaining
agents are represented as isolated islands.
\end{comment}

The diameters of the various layers support these observations. The largest shortest distance between any two agents is approximately four to five edges for the credit, cross-securities, and funding layers, while it is only two edges for the overlapping portfolio layer. This suggests that fire sales by agents can propagate much more rapidly through the network in comparison to default cascading effects observed in the other layers.

\begin{figure}[!htbp]
  \centering
  \includegraphics[width=0.95\textwidth]{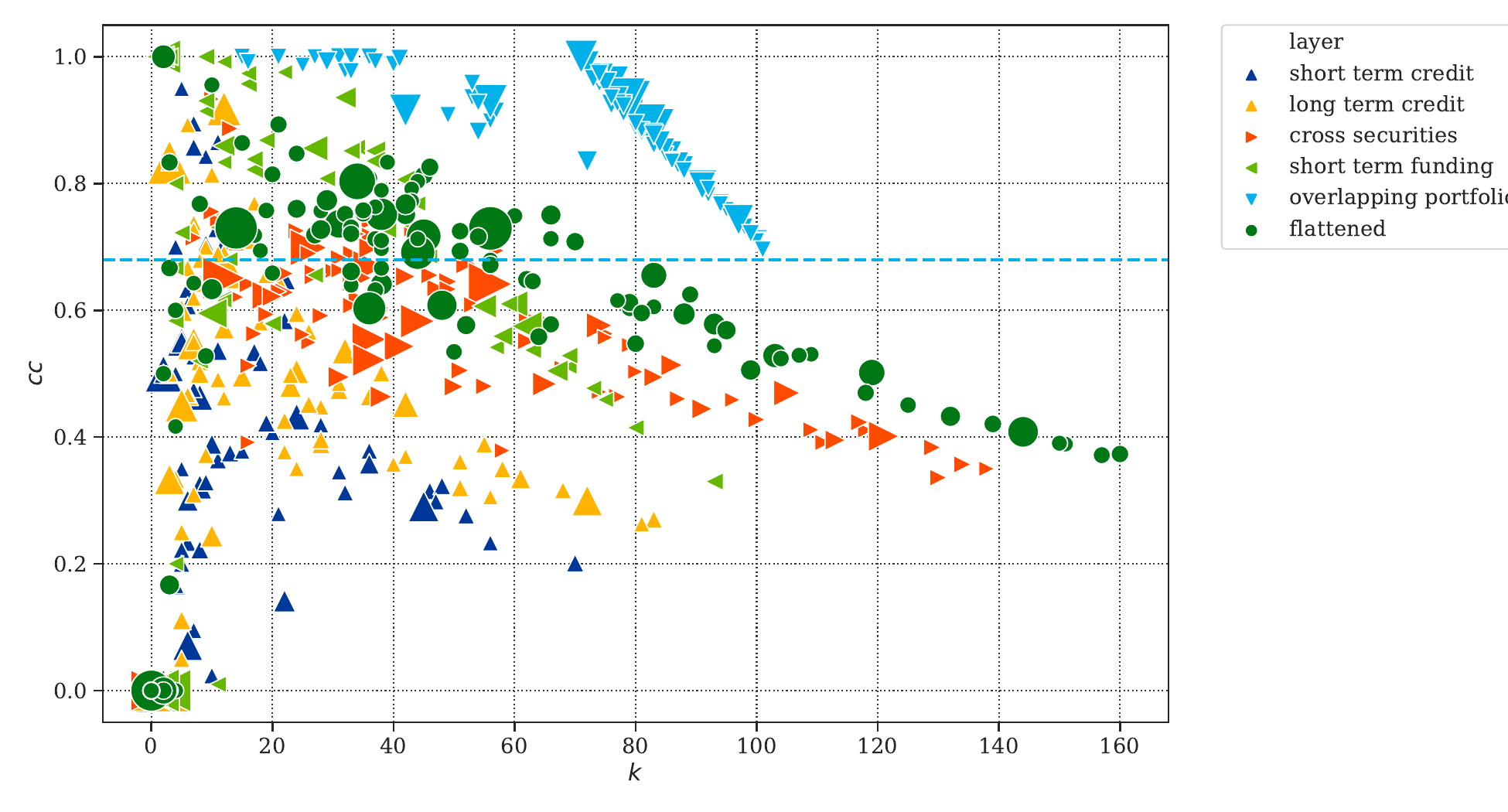}
  \caption{\footnotesize Relationship between the degree ($k$) and clustering coefficient ($cc$) of banking groups within the multi-layer network. Each layer is color-coded, and the size of the markers is proportional to the total assets of the banking groups. The figure has been created using~\cite{Hunter:2007}.}\label{fig1}
\end{figure}

Turning to the clustering of nodes, Table~\ref{tab1} displays the average clustering coefficients across the markets. The clustering coefficient ranges from 29\% (short-term credit layer) to 51\% (cross-securities layer).\footnote{It is worth mentioning that the clustering coefficient of the overlapping portfolio layer is substantially higher due to its undirected nature. Assuming a directed graph, the clustering coefficient aligns more closely with other layers at 39\%.} These values are consistent with empirical observations in other interbank markets (\cite{Sor2007}; \cite{bargigli2015multiplex}). In conjunction with the low diameters, all layers demonstrate potential small-world properties as defined by \cite{Watts1998}. To explore this further, we follow the approach of \cite{ContMoussaSantos} and analyze the relationship between degree distributions and clustering coefficients across the layers. Figure~\ref{fig1} illustrates the results. Despite some noise, an overall negative relationship between these metrics is apparent. This suggests that highly connected agents tend to interact with counterparties who have limited interactions with each other, typically smaller entities. This observation indicates the presence of smaller-sized hubs connecting diverse market participants. In contrast, agents with lower degrees are interconnected with counterparties who are densely connected with each other, encompassing varying entity sizes, including the largest agents. This configuration hints at the existence of stable and robust relationships among certain agents. Notably, only in the overlapping portfolio layer can we confirm that the clustering coefficient is bounded away from zero. Thus, unlike other layers and in conjunction with the graph statistics, the overlapping portfolio layer stands out as a potential small-world graph.

\subsection{Scale-free properties}

% This part can potentially be moved to introduction/related work
Theoretical literature on interbank networks typically assume so-called scale-free networks when simulating plausible interbank networks (e.g.; \cite{GonzlezAvella2016}, \cite{Montagna2016} and \cite{Soramki2013}) These types of networks commonly rely on specific statistical properties of the tails of the degree distributions underlying the networks. Technically, scale-free networks are loosely described by a probability density function which can be modelled by a power law functional relation, at least asymptotically. In this section, we examine the multi-layer network to determine whether these assumptions in literature are in line with empirical findings and how these findings might change across the different layers in the network. The premise about the potential of scale-free networks to capture the underlying network topology of interbank markets holds great importance for economic policy~\cite{montagna2016multi}. Contagion models often rely on scale-free topologies to simulate an interbank network (e.g. \cite{Soramki2013}), which has direct consequences on various policy implications \cite{battiston2012debtrank}. %Thus, validating assumptions like scale-free properties is crucial to increase the certitude of stress test scenario’s that rely upon them.

\begin{comment}
To examine these properties, we conduct the following exercise: first we
disaggregate the multilayer network from banking group level to the
entity level, in order to look at interactions between banks from a more
granular scope. During the disaggregation, we make sure to filter away
any intra-group exposures, to keep our focus on transactions between
entities from different banking groups. Next, we gather a set of
candidate statistical distributions that are known to fit heavy-tailed
data and fit them on our empirical data. Afterwards, to further pin down the
number of candidates, we leverage a likelihood ratio test to determine
which candidate fits the data the best from a statistical point of view.
Lastly, we perform a semi-parametric bootstrapping approach a la mode de
(Clauset, Shalizi, \& Newman, 2007), to determine the statistical
significance of the goodness-of-fit of the main candidate distribution.
\end{comment}

Our analysis focuses on the distribution of the weighted in-degree of
the entities in the multi-layer network. The weighted in-degree of an
entity $K_{weighted}^{in}(i)$, represents the total of
exposures connected to that entity $n_i$. E.g., for the long-term credit
market $K_{weighted}^{in}(i)$ is equal to the total
amount that $n_i$ has borrowed from various creditors belonging to
banking groups where $n_i$ is not part of. A similar analysis can be
set-up for the weighted out-degree $K_{weighted}^{out}(i)$.

We consider the set of statistical distributions depicted in
Table~\ref{tab2} as the candidates to fit $K_{weighted}^{in}(i)$.~\footnote{The third column of Table~\ref{tab2} shows the normalization constant \(C\)
such that \(\sum_{x = x_{xmin}}^{\infty\ \ }{Cf(x)} = 1\).} The power law distribution is of main interest as this
gives most evidence of the existence of scale-free networks and is
usually embedded in the generating process when simulating such
networks~\cite{Clauset2009}. We also consider the truncated power law which behaves similar
to the power law but converges towards an exponential distribution as we
move deeper into the tails. Next, we consider the lognormal distribution
due to its flexibility of capturing heavy tails~\cite{van2014finding}. Lastly, we consider the
exponential distribution which is, per definition, the minimum
alternative candidate for evaluating the existence of heavy-tails, as
heavy-tailed distributions are usually defined as not
exponentially bounded \cite{Markovich2018}.

\begin{threeparttable}
\setlength{\tabcolsep}{10pt}
\begin{longtable}[]{@{}lll@{}}
\caption{Statistical distributions to approximate empirical data}\label{tab2} \\
\toprule
Name & functional form \(\mathbf{f(x)}\) & normalization constant \(\mathbf{C}\)\tabularnewline
\midrule
\endhead
Power law & \(x^{- \alpha}\) &
\(\dfrac{1}{\zeta\!\left( \alpha,\ x_{\min} \right)}\)\tabularnewline
Truncated power law & \(x^{- \alpha}e^{- \lambda x}\) &
\(\dfrac{\lambda^{1 - \alpha}}{\Gamma\!\left( 1 - \alpha,\lambda x_{\min} \right)}\)\tabularnewline
Lognormal &
\(\dfrac{1}{x}\exp\!\left( - \dfrac{\left( \ln x - \mu \right)^{2}}{2\sigma^{2}} \right)\) &
\(\Bigl[\sigma\sqrt{2\pi}\cdot \tfrac{1}{2}\,\mathrm{erfc}\!\Bigl(\dfrac{\ln x_{\min}-\mu}{\sqrt{2}\sigma}\Bigr)\Bigr]^{-1}\)\tabularnewline
Exponential & \(e^{- \lambda x}\) &
\((1 - e^{- \lambda})e^{\lambda x_{\min}}\)\tabularnewline
\bottomrule
\end{longtable}
\end{threeparttable} \\

\begin{figure}[!htbp]
  \centering
  \includegraphics[width=0.95\textwidth]{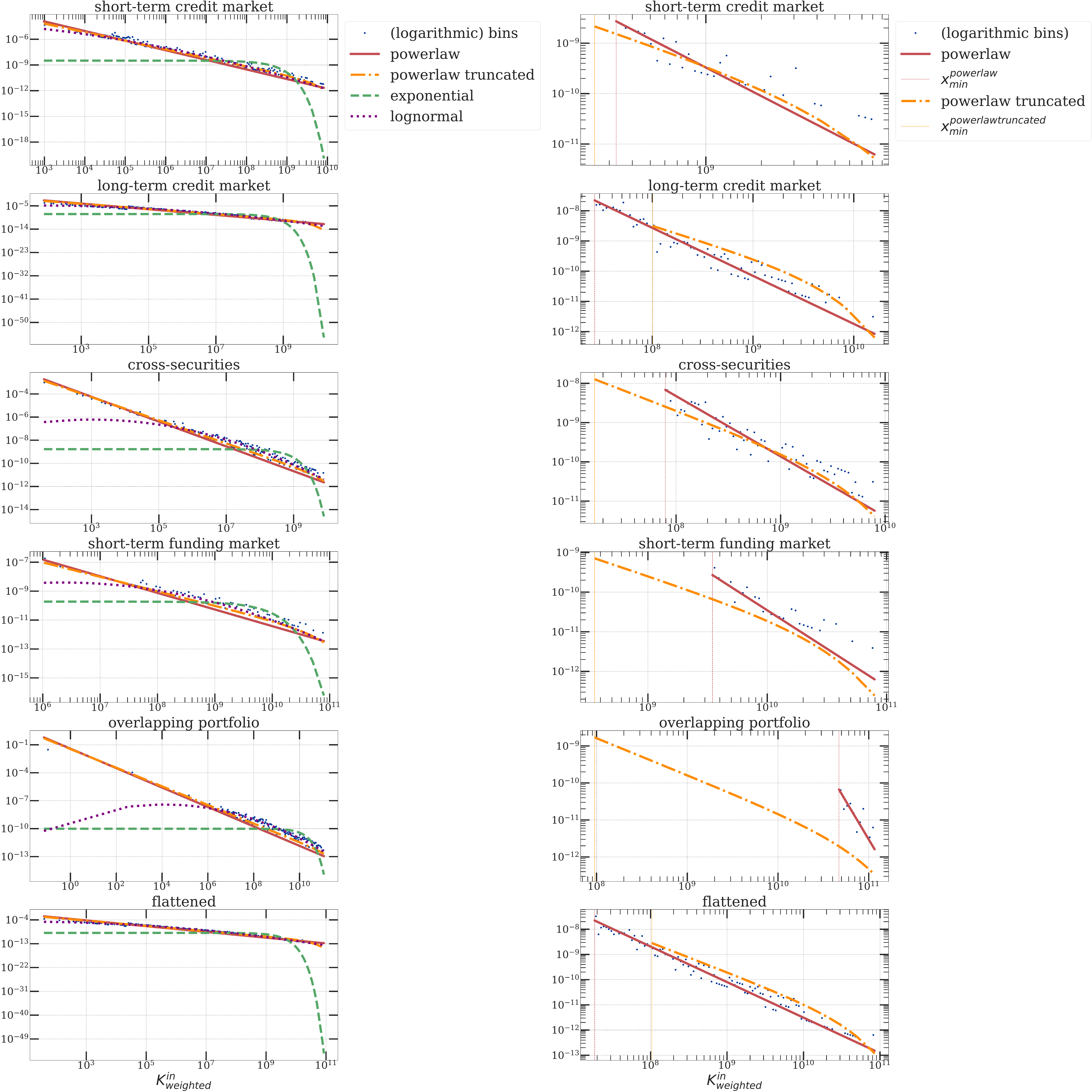}
  \caption{ \footnotesize The figure shows the empirical probability density functions (PDFs) over the entire data-range (left column) as well as their tail-only estimation (second column). Empirical PDFs are estimated using logarithmic binning in combination with the normalizing scheme of \cite{Milojevi2010}. Empirical PDFs are depicted by the centre of these logarithmic bins as blue markers presented on logarithmic spaced axes (log-log axes). Fitted candidate distributions are presented as colour coded lines. Candidate distributions are fitted by following the methodology depicted in~\cite{Clauset2009}, and utilizing the Python libraries offered by~\cite{2020SciPy-NMeth} and~\cite{Alstott_2014}. The figure has been created using~\cite{Hunter:2007}.}\label{fig2}
\end{figure}

Figure~\ref{fig2} plots the empirical probability density functions (PDFs, blue markers).
For each of the different layers in the multi-layer network we fit the candidate distributions using the method of maximum likelihood. In other words, we utilize the respective Maximum Likelihood Estimators (MLEs) of the candidates to find the best fitting parameter(s) to describe the empirical observations. The left column of Figure~\ref{fig2} overlays the fitted candidate distributions when using all of the empirical data. The right column of Figure~\ref{fig2} represents the same information as the first column, but here we use the optimization scheme of \cite{Clauset2009} to fit a (truncated) power law to only the (estimated) tail of the empirical distributions.

In all layers, we observe weighted in-degree exposures spanning several orders of magnitude, indicating a heavy-tailed distribution. In the short-term credit layer, the empirical probability density function (PDF) appears relatively linear, with most candidate distributions fitting the data well except for the exponential distribution. Focusing on the right column of Figure~\ref{fig2}, both the truncated and the traditional power law models estimate the start of the tail distribution, denoted as \(x_{min}\), at approximately \(30 \times 10^7\) EUR, roughly the \(80^{th}\) percentile. The long-term credit layer and the flattened layer exhibit similar characteristics. The exponential distribution misfits the data from the short-term credit, cross-securities, short-term funding, and overlapping portfolio layers, whereas the truncated and the traditional power law models better capture the entire data range. Notably, tail estimation for the short-term funding layer significantly differs between the truncated and traditional power law, with the power law estimating a narrower tail. Tail estimations for the overlapping portfolio layer appear problematic, with estimated lower bounds of the tails nearing the upper edge of the distribution. This suggests that no discernible difference exists between the bulk data and its tail. For such estimations the identified tail does not
hold a sufficient amount of data points to perform a meaningful fit.

To quantitatively determine which candidate best fits the empirical data, we leverage the following pairwise likelihood ratio test \(\mathcal{R}\):

\[{\mathcal{R}(p_{1},p_{2})}{= \ \sum_{i = 1}^{N}{\bigl[\ln{p_{1}\left( x_{i} \right) - \ln{p_{2}\left( x_{i} \right)}}\bigr]}}.\]

With \(p_{1}(x)\) and \(p_{2}(x)\) the PDFs of the theoretical
distributions and \(N\) the total number of empirical observations. A positive value indicates that \(p_{1}(x)\) fits the data better than \(p_{2}(x)\). To
determine the best fitting distribution out of all candidates, we employ a similar methodology as \cite{Vandermarliere2015}, where we compute the following scores:
$$
g(p_k) = \sum_{l\neq k}{\bigl[ \mathcal{R}(p_k,p_l) \text{ if p-value} \leq 0.05 \bigr]}
$$
With $k$ and $l$ one of the candidate
distributions. The candidate with the highest score is deemed the most
suitable theoretical distribution to explain the observed data.

\begin{table}[t]
\centering
\caption{Best-fitting distributions (bulk + tail)}
\label{tab:best_fit_bulk_tail}
\small
\setlength{\tabcolsep}{5pt}

\begin{tabular}{l l r l r}
\toprule
Layer & Best fit & Score & Runner-up & LR test ($\mathcal{R}$) \\
\midrule
Short-term credit      & Truncated power law &  27.47 & Lognormal &  -0.53$^{**}$ \\
Long-term credit       & Truncated power law &  41.43 & Lognormal & -17.47$^{***}$ \\
Cross-securities       & Truncated power law &  57.62 & Lognormal &  -7.79$^{***}$ \\
Repo market            & Truncated power law &  18.93 & Lognormal &  -2.48$^{*}$ \\
Overlapping portfolio  & Truncated power law & 129.80 & Lognormal & -11.75$^{***}$ \\
Flattened network      & Truncated power law &  48.81 & Lognormal & -21.73$^{***}$ \\
\bottomrule
\end{tabular}

\vspace{2pt}
\begin{minipage}{0.98\linewidth}
\footnotesize
\emph{Notes:} The last column reports the likelihood-ratio statistic $\mathcal{R}$ comparing the best fit against the runner-up. $^{***}$, $^{**}$, and $^{*}$ denote significance at the 1\%, 5\%, and 10\% levels, respectively, following \cite{Vuong1989}.
\end{minipage}
\end{table}

\subsection{Semi-parametric bootstrapped goodness-of-fit}\label{sec:boot_gof}

Likelihood-ratio tests only indicate which candidate distribution provides the \emph{relatively} better fit
to the empirical data; they do not imply that the best-performing candidate is an \emph{adequate} description
of the data-generating process. In fact, for any empirical sample one can typically fit some theoretical
distribution. To assess absolute fit, \cite{Clauset2009} propose a semi-parametric bootstrap procedure
for goodness-of-fit testing of heavy-tailed distributions, which also applies when the tail is fitted above an
estimated cut-off value $\hat{x}_{\min}$.

\begin{figure}[tb]
  \centering
  \includegraphics[width=\linewidth]{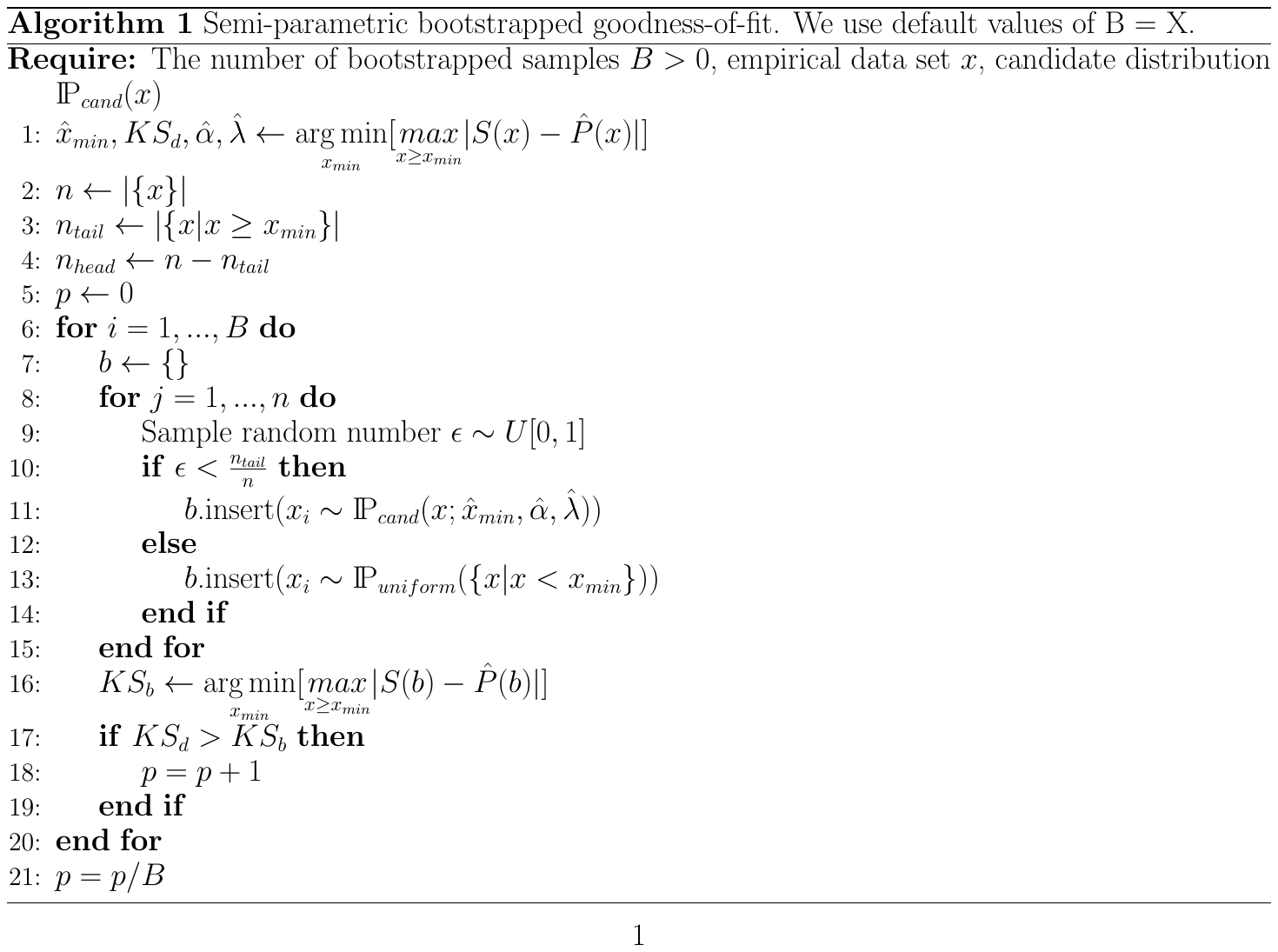}
  \caption{The bootstrapping algorithm follows the semi-parametric procedure of
\cite{Clauset2009} to assess the goodness-of-fit of heavy-tailed distributions. The approach can also be used when estimating a tail distribution with an estimated cut-off value $\hat{x}_{\min}$.}
  \label{box:boot_gof}
\end{figure}

\begin{figure}[!htbp]
  \centering
  \includegraphics[width=0.9\textwidth]{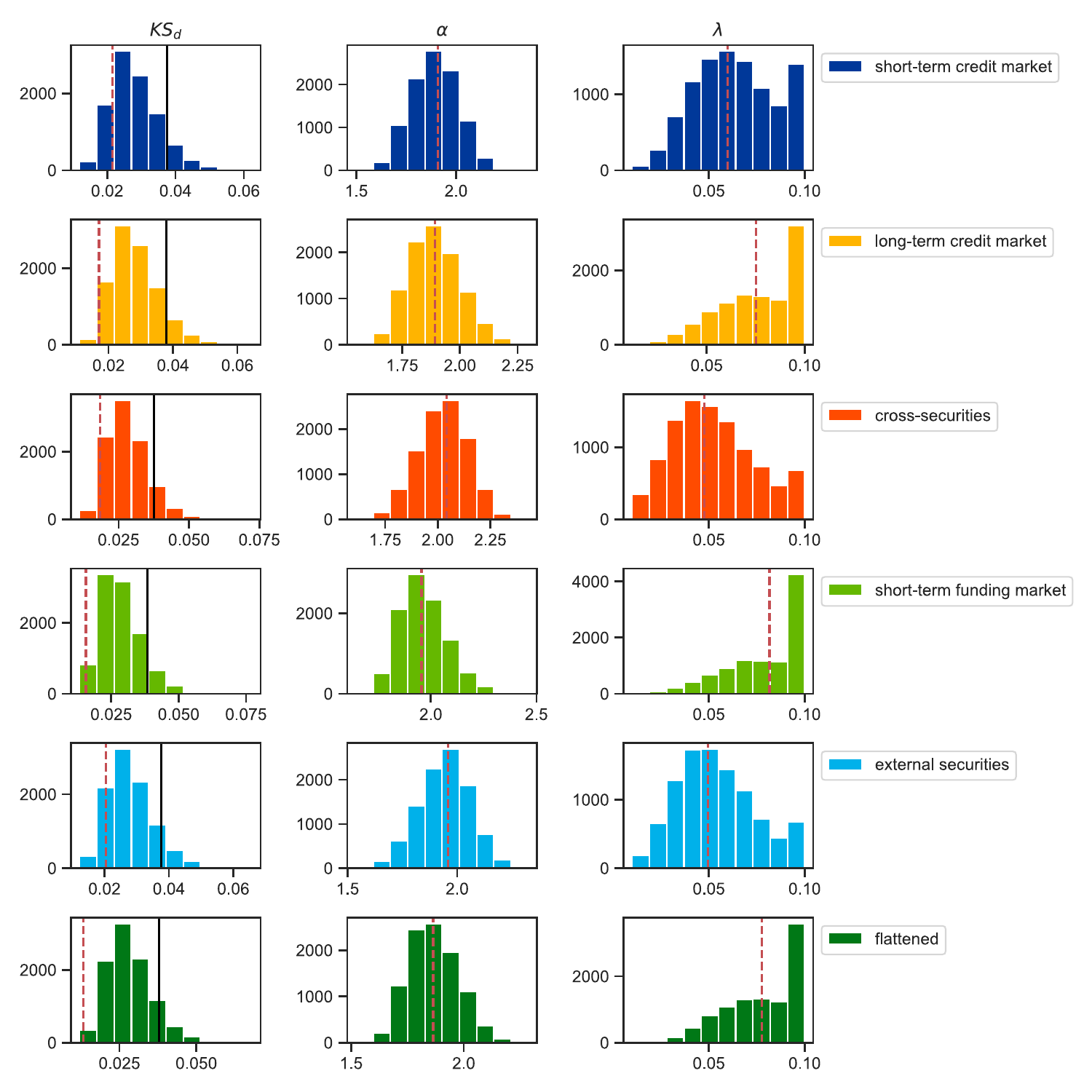}
  \caption{Semi-parametric bootstrapping results.
  Rows correspond to layers. The first column reports bootstrap Kolmogorov--Smirnov distances; the second column
  shows bootstrap lower-bound estimates $\hat{x}_{\min}$ (when performing tail-only fits); the remaining columns
  report bootstrap parameter estimates (e.g., $\alpha$ and $\lambda$) for the truncated power law. The vertical
  red dashed line marks the non-bootstrapped estimate from the empirical data, and the solid black line indicates
  the 90th percentile of the bootstrap distribution.}
  \label{fig:boot_results}
\end{figure}

The bootstrap procedure is summarized in Figure~\ref{box:boot_gof}, and the results for the full empirical range
(bulk + tail) are reported in Figure~\ref{fig:boot_results}. Guided by the likelihood-ratio outcomes, we apply
this scheme to test whether a truncated power law provides an adequate fit to the empirically observed degree
distributions in each layer. The null hypothesis, $H_0$, is that the data are generated by a truncated power law.
More precisely, $H_0$ states that the (positive) Kolmogorov--Smirnov distance between the empirical distribution
and the fitted truncated power law can be attributed to statistical fluctuations under the truncated power-law
data-generating process. We simulate $B$ bootstrap samples from the fitted model, re-estimate the parameters on
each synthetic sample, and recompute the Kolmogorov--Smirnov distance. If the empirical distance is large relative
to the bootstrap distribution (i.e., many bootstrap distances are smaller than the empirical one), then $H_0$ is
unlikely and should be rejected. We use a conservative confidence level of 10\%\footnote{In this setting, higher
confidence levels make it easier to confirm a truncated power-law data-generating process, whereas lower levels
make confirmation more stringent.} to decide on rejection.

As shown in Figure~\ref{fig:boot_results}, for all layers we fail to reject $H_0$, and thus find evidence that the
(weighted) degree distributions are consistent with a truncated power law. The empirical Kolmogorov--Smirnov
distances are sufficiently small relative to their bootstrap counterparts to be plausibly explained by sampling
variability, suggesting that these markets are governed by a truncated power-law data-generating mechanism. The
layers are also broadly consistent in terms of parameter estimates, with $\alpha \in [1.75, 2.25]$ and
$\lambda \in [0.05, 0.075]$, in line with values reported by \cite{Vandermarliere2015} and \cite{ContMoussaSantos}.

Despite pronounced topological differences across layers, the weighted in-degree distributions display remarkable
regularity. By contrast, when restricting attention to tail-only fitting, confirming a truncated power law becomes
substantially more difficult. In unreported results we reject $H_0$ for the long-term credit, cross-securities,
and flattened layers, indicating alternative data-generating processes; however, these tail-only conclusions are
weaker given the limited number of observations in the extreme tail.

The conclusions above depend on the network characteristic under study. In particular, when examining unweighted
degree distributions (both in- and out-degree), we observe substantially more heterogeneity across layers, with
evidence consistent with distinct data-generating processes. Researchers and practitioners should therefore be
careful when simulating networks to represent specific segments of the interbank system and should ensure that the
chosen representation reflects the underlying market relationships they aim to proxy.

\subsection{Centrality measures}

% can be moved to intro/related work
Centrality measures assess the centrality or degree of connection of a node within a specific layer of a multi-layer network. Typically, these metrics are derived from factors such as the number of connections, the significance of these connections, and the nodes they link, representing a node's importance in the overall network structure \cite{Borgatti2006}. While these measures may not be identical, they bear a resemblance to the concept of systemic importance extensively studied in finance and economics. Quantifying the primary drivers of systemic risk and attributing risk to individual contributors pose significant challenges \cite{Tarashev2009}. Utilizing graph-based models for interbank markets and leveraging centrality measures present a promising approach to address some of these challenges. As emphasized by \cite{huser2015too}, theoretical literature on interbank networks offers a coherent method for studying interconnections, contagion processes, and systemic risk, albeit with certain limitations. Nevertheless, centrality measures can offer supplementary insights into the systemic significance of banks, surpassing traditional financial stability metrics.

In general, centrality measures yield the flexibility to introduce a ranking of nodes according to their tenor within a layer, allowing one to discover the most pivotal of nodes within the system~\cite{bloch2023centrality}. Furthermore, a comparison of a centrality measure of the same node across different layers provides insights on the differences in the topology of layers. If a node is considered very central in one layer, but less central in an another, then this underlines the different semantics of the layers.

\begin{figure}[!htbp]
  \centering
  \includegraphics[width=0.95\textwidth]{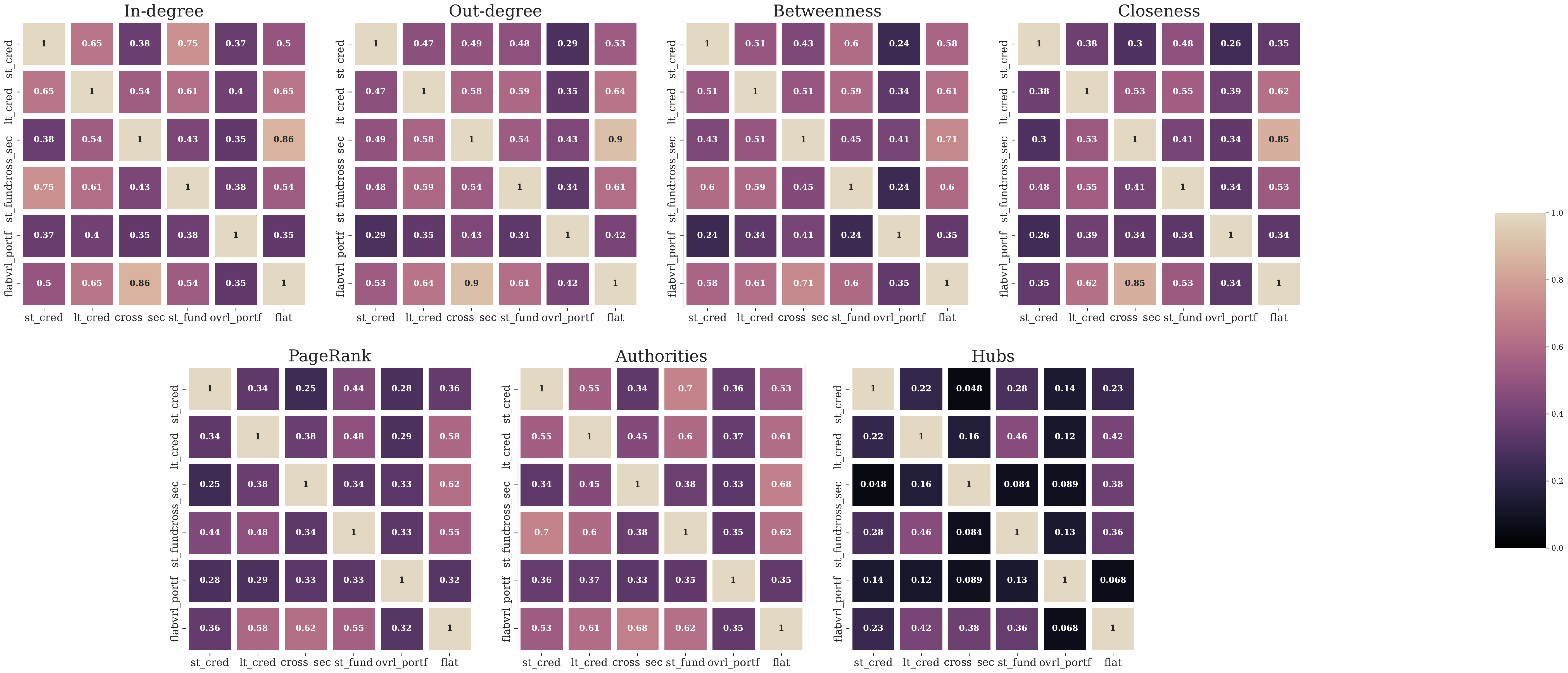}
  \caption{\footnotesize Each matrix represents the set of pairwise Kendall rank correlations based on a certain centrality measure. Names of the layers are abbreviated: short-term credit layer (st\_cred), long-term credit layer (lt\_cred), cross-securities layer (cross\_sec), short-term funding layer (st\_fund), overlapping portfolio layer (ovrl\_portfl) and the flattened layer (flat). The figure has been created using~\cite{Hunter:2007}.}
  \label{fig3}
\end{figure}

We examine the cross-correlation patterns of various centrality measures across the layers. Figure~\ref{fig3} shows the Kendall Rank correlation matrices. The Kendall Rank correlation is a statistic used to measure the ordinal association between two measured quantities~\cite{kendall1970rank}. The Kendall correlation between two evaluators will be high (with an upper bound of 1) when observations have a similar ranking between the two evaluators and low (lower bound of -1) when observations have a dissimilar ranking between the evaluators. For our cause, we appoint the different layers as the evaluators, and analyze their agreement of which nodes are most central. The first observation from Figure~\ref{fig3} is that strictly positive correlations are observed for all centrality measures, hinting at some agreement between the layers. On a closer look, we see that most correlation values are between 30-60\%, with an exception of the ranking correlation between the cross-securities market and the flattened network, which exceed the 80\% mark for most centrality measures. Another exception is the Hubs measure for which low agreement is observed between most layers.

\begin{figure}[!htbp]
  \centering
  \includegraphics[width=0.95\textwidth]{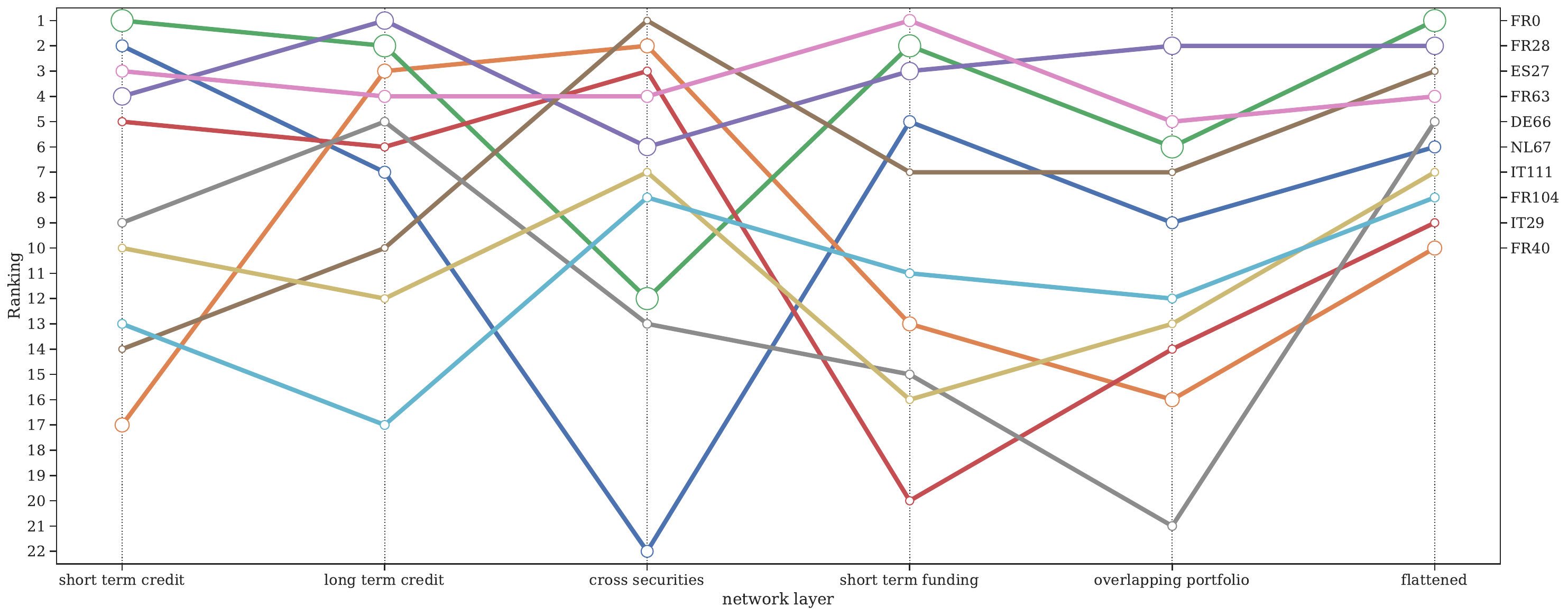}
  \caption{\footnotesize The ranking of the 10 largest banking groups by the PageRank algorithm $C^{PR}(i)$ across different layers of the multi-layer network. Banking groups are color-coded and the size of the markers is proportional to the total assets of the banking groups. The figure has been created using~\cite{Hunter:2007}.}
  \label{fig4}
\end{figure}

To illustrate, we focus on node centrality assessed through the PageRank algorithm \(C^{PR}(i)\). Figure~\ref{fig4} showcases the top 10 banking groups in various network layers, determined by their PageRank centrality scores. In each layer, the PageRank algorithm was applied, generating rankings to identify the most central banking groups within that specific layer. The results reveal clear heterogeneity between layers; certain nodes are highly central in one layer but not nearly as significant in another. For instance, the largest banking group FR0 (depicted in green) ranks as the most central node in the short-term credit market but doesn't even make it to the top 10 central nodes in the cross-securities market. Similar disparities are observed for other nodes in the network.

In accordance with the Kendall Rank correlation matrices, we find a level of centrality persistence across layers, indicating that nodes ranked highly within one layer are more likely to be within the top 20\% percentile in other layers. However, this persistence is not precise; within this top 20\%, there is considerable heterogeneity, with nodes that are top-ranked in one layer not necessarily maintaining top positions in other layers.

\section{Concluding remarks} % .5 pages

In this paper, we demonstrate that recent Euro Area granular data collections serve as a robust foundation for constructing and analyzing multi-layer networks. The multi-layer network framework described in our work provides a natural and effective means to represent granular data, enabling a comprehensive understanding of the topological structure of the EA financial interbank system. Our analysis sheds light on the intricate nature of contemporary interbank markets, offering insights into a wide spectrum of network layers and their diverse topological properties. By benchmarking the real-world data-derived network against common theoretical assumptions, our findings underscore the necessity for researchers and practitioners to exercise caution when simulating networks to target specific segments of the interbank network. It is important to acknowledge and incorporate the intricacies inherent in the financial markets being modeled. That is, some assumptions (e.g., small-world properties or power-law behaviour) are valid for only a certain type of financial relationship between banking groups.

\printbibliography[heading=bibintoc, title=\ebibname]

\end{document}